\documentclass[11pt,twoside]{article}

\usepackage{asp2004}
\usepackage{epsf}
\usepackage{psfig}
\usepackage{lscape}
\usepackage{graphicx}

\newcommand{\Tef}{T$_{\rm eff}$~}
\newcommand{\Vm}{$V_{\rm macro}$~}
\newcommand{\Vt}{$V_{\rm t}$~}
\newcommand{\Rs}{R$_\odot$~}

\newcommand{\Vg}{$V_{\rm g}$~}

\newcommand{\HHO}{H$_2$O~}
\newcommand{\vmon}{V838 Mon~}

\begin{document}
\title{Modelling the spectrum and SED of V838 Mon}
\author{Ya.Pavlenko, B. Kaminsky, Yu. Lyubchik, La. Yakovina}
\affil{Main Astronomical observatory of NASU, 27 Zabolotnoho, Kyiv-127, 03680
Ukraine}

\begin{abstract}
We model the spectra and spectral energy distribution of V838 Mon
which were observed in February, March, and November, 2002. Theoretical
spectra are calculated using the classical model atmospheres
taking into account absorption of atomic and molecular lines.
By fitting the observed spectra we determine the physical
parameters of the atmosphere of V838 Mon. These parameters
are determined to be Teff = 5330 $\pm$ 300
K, 5540 $\pm$ 270 K, 4960 $\pm$ 190 K, and 2000 $\pm$ 200 K for
February 25, March 2, March 26, and November 6, respectively.
\end{abstract}

\subsection{Introduction}

The cause of the eruption of \vmon\ and the nature of its progenitor
are unclear. Desidera et al. (2002) reported a
faint hot continuum at short wavelengths, recently confirmed by
Munari et al. (2005) and identified as a B3V companion.

\section{Procedure}

\subsection{Observational data}

\begin{figure*}
\begin{center}
\includegraphics [width=65mm, angle=00]{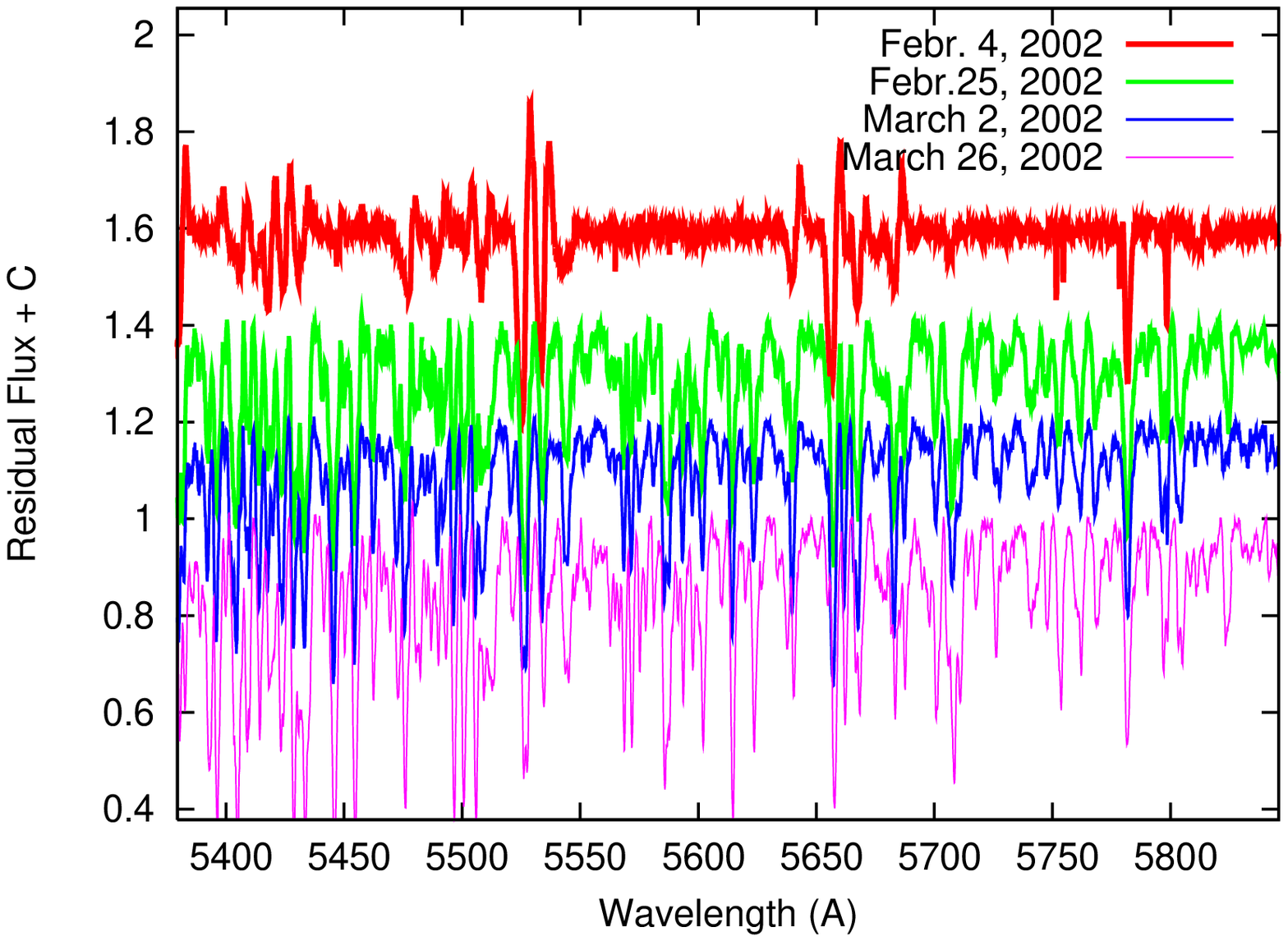}
\includegraphics [width=65mm, angle=00]{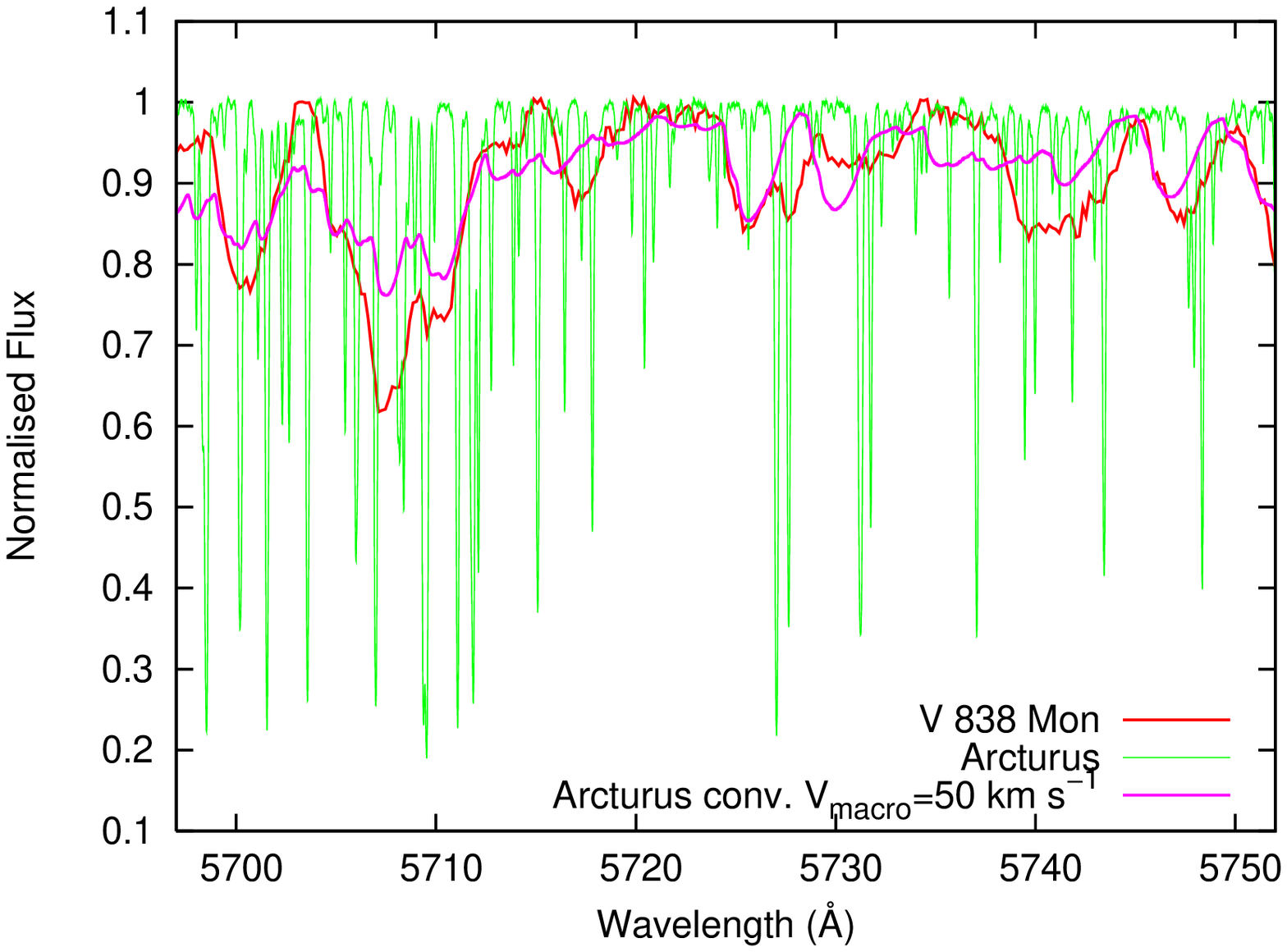}
\end{center}
\caption[]{\label{__obs} Left: a comparison of observed spectra of
\vmon\ for different dates in Febr. -- March, 2002. Right:
comparison of the spectrum of V838 Mon and that of  Arcturus
(Hinkle et al. 2000), convolved with macroturbulent profile \Vm
= 50 km s$^{-1}$ See Kaminsky \& Pavlenko (2002) for more
details.}
\end{figure*}

Some spectra obtained on February -- March, 2002 are shown in Fig. \ref{__obs}.

The first spectrum of V838 Mon taken on Feb. 4 is shown in Fig. \ref{__obs}.
It was obtained by V.Klochkova 
at the 6 m telescope (SAO, Russia). Details of the data reduction can be found in 
Kipper et al. (2004).

On February 25 and March 26, 2002
spectra (R $\sim$ 18000) of V838 Mon were obtained
with the Echelle+CCD spectrograph on the 1.82m telescope operated
by Osservatorio Astronomico di Padova on Mount Ekar (Asiago).

Spectra  on March 2 were
obtained with the echelle fibre-fed spectrograph on the
1.9-m SAAO telescope kindly provided for us by Lisa Crause (see
Crause et al. 2003 for details).

The November 6, 2002 spectrum was obtained with the Kast
spectograph on the Cassegrain focus of the Shane 3-m telescope at Lick
Observatory. Full details of the observing and data-reduction processes
are given in Rushton et al. (2005) and are not repeated here.

\subsection{Computation of theoretical spectra}

Through February to March 2002, the spectrum of V838 Mon 
looks like a normal spectrum of a
supergiant of spectral class K -- G. The optical spectra of V838Mon over these dates
are governed by absorption by the neutral atoms. A comparison with a synthetic
spectrum of Acturus, with lines broadened by macroturbulent velocities of 50 km/s,
shows good agreement (Fig. \ref{__obs}).

To model these spectra we used classical model atmospheres from the Kurucz (1993)
grid.
Computations of the
synthetic spectra were carried out using the program WITA6
Pavlenko (2000) assuming LTE and hydrostatic equilibrium for a
one-dimensional model atmosphere without sources and sinks of
energy.The equations of ionization-dissociation equilibrium were solved
for media consisting of atoms, ions and molecules.
Computations were carried out for the VALD (Kupka et al. 1999) line list.

Up until November 2002 both effective temperature and luminosity of V838 Mon drop
significantly with time (see also Tylenda 2005).
The spectrum of V838 Mon contains strong molecular features. To compute
theoretical spectral energy distributions (SEDs) we used
a number of model atmospheres, with
\Tef = 2000 --- 2200~K, log g = 0, 0.5 from the NextGen grid of
Hauschildt et al. (1999).

In addition to VALD,
molecular line data were taken from a variety of sources: \\
-- the TiO line lists of Plez (1998). \\
-- CN lines from CDROM 18 (Kurucz 1993); \\
-- CrH and FeH lines from Burrows (2002) and
      Dulick et al. (2003) respectively \\
-- lines of H$_2^{16}$O from the BT2 database
      (Barber et al. 2006) \\
-- absorption by VO, and by a few molecules of (in the case of
      \vmon) lesser importance, was computed in the JOLA
      approximation (see Pavlenko et al. 2000).

The relative importance of the different opacities contributing to
our synthetic spectra is shown in Fig. \ref{__obs}.

\begin{figure*}
\includegraphics [width=65mm]{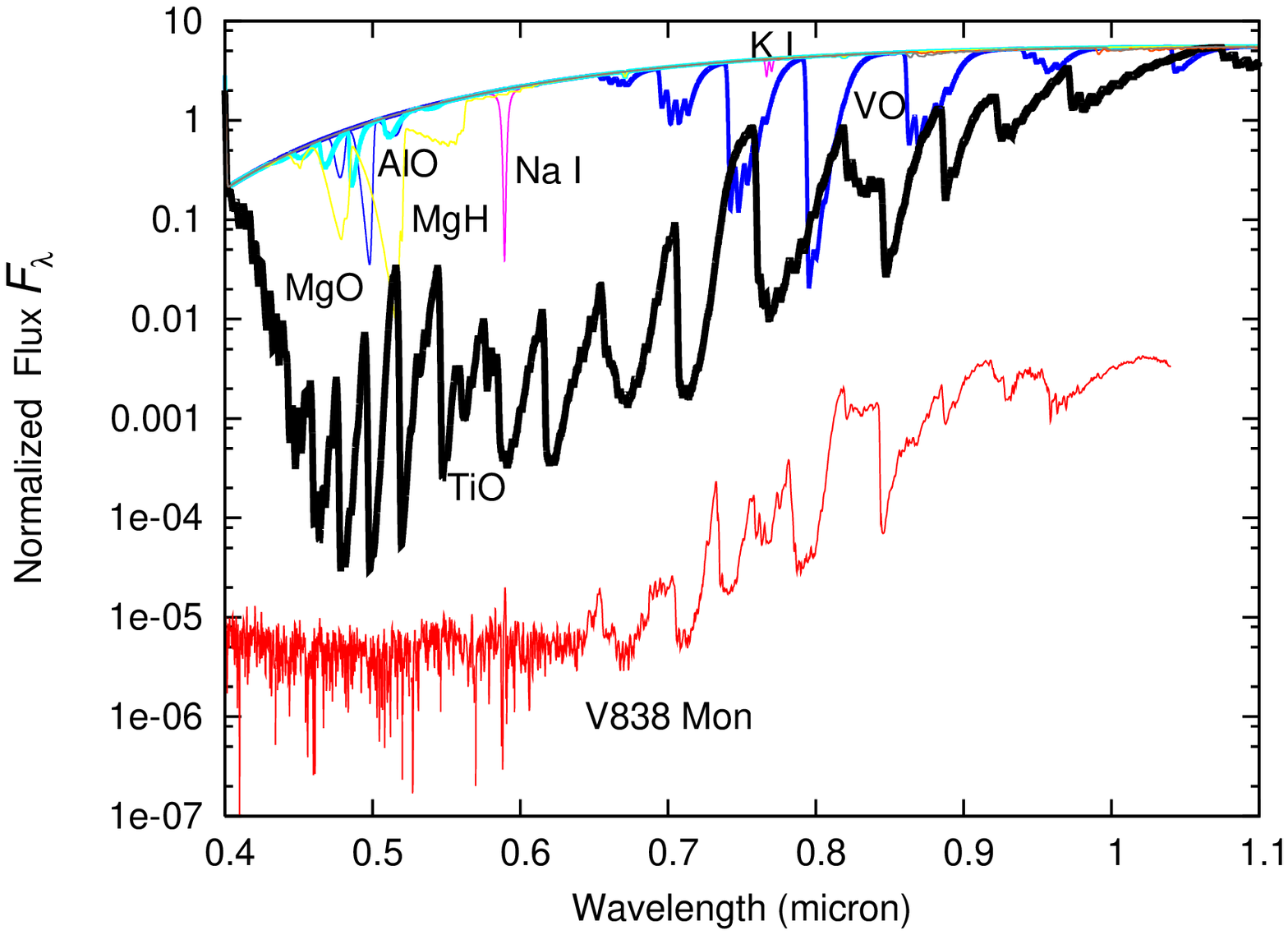}
\includegraphics [width=65mm]{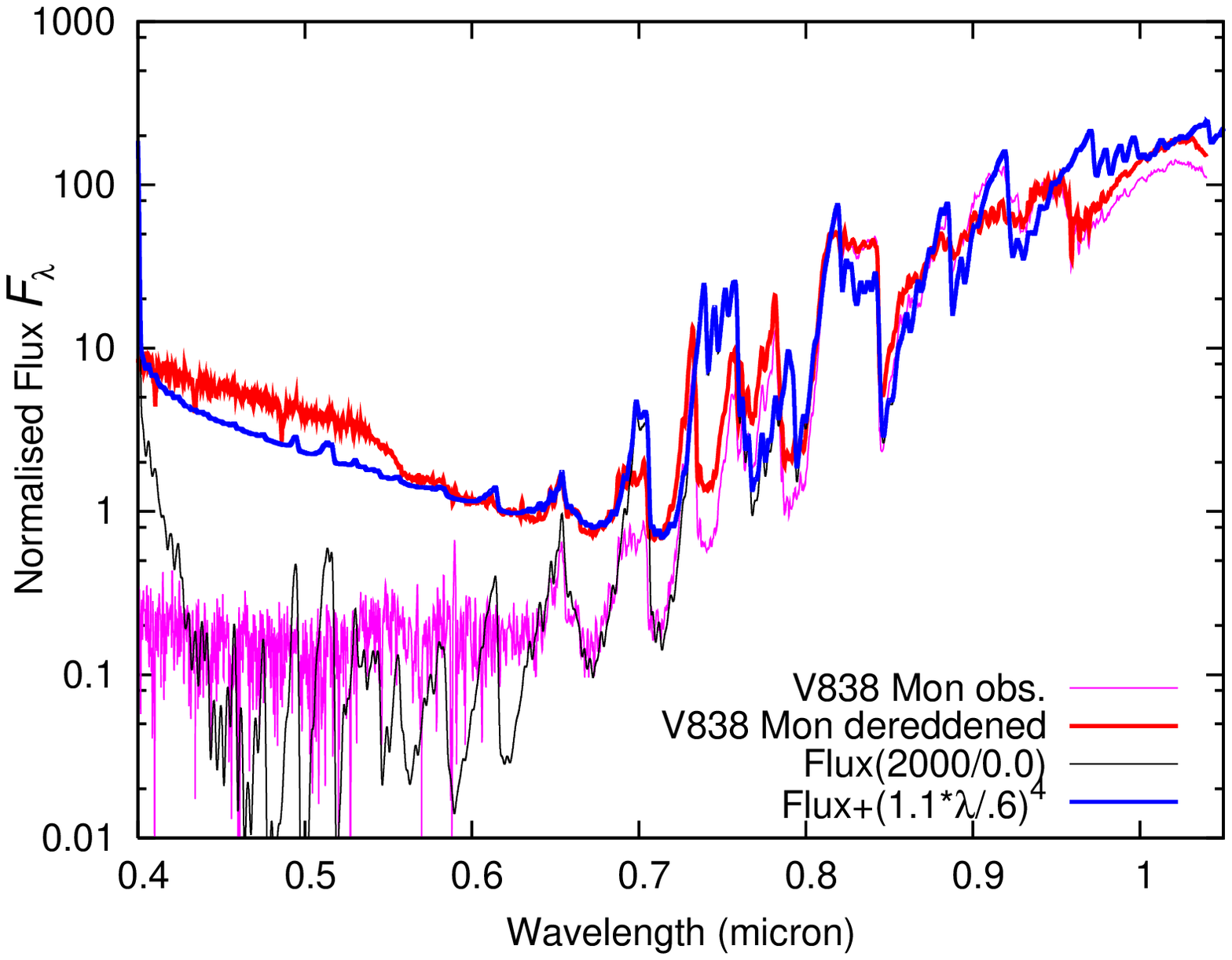}
\caption{\label{__nov}Left: the contribution of different molecules to the
formation of the spectrum of \vmon. Right: our fit 
to the observed \vmon spectrum on November 6, 2002 (Pavlenko et al. 2006).}

\end{figure*}

\section{Results}

\subsection{V838 Mon in February -- March, 2002}

It is worth noting a few results (see Kaminsky \& Pavlenko
(2005) for more details of our procedure and results ):

For February 25 we obtained \Tef = 5330 $\pm 300$ K, $\log
N(\mbox{Fe}) = -4.7 \pm 0.14$ dex and \Vt = 13. $\pm$ 2.8 km
s$^{-1}$.

For March 2 -- \Tef = 5540 $\pm$ 190 K,
$\log N(\mbox{Fe}) = -4.75 \pm 0.14$ dex, \Vt = 13.3 $\pm$ 3.2 km
s$^{-1}$.

For March 26 -- \Tef = 4960 $\pm$ 270
K, $\log N(\mbox{Fe}) = -4.68 \pm 0.11$ dex, \Vt = 12.5 $\pm$ 1.7
km s$^{-1}$.

To account for processes of broadening of spectral lines in spectrum of
V838 Mon we used parameter \Vg. In our case \Vg  represents the
macroturbulent
velocity.
We obtained \Vg = 54 $\pm$ 3, 47 $\pm$ 3 and 42 $\pm$ 5 km
s$^{-1}$ for February 25, March 2 and March 26, respectively.

Finally, we obtained changes of the radial velocity V$_{\rm radial}$ = $-$76 $\pm$ 3, $-$70
$\pm$ 3 and $-$65 $\pm$ 3 km s$^{-1}$ for February 25, March 2 and
March 26, respectively, it appears possible that there is a reduction in
the expansion velocity of the envelope.


\subsection{V838 Mon in November, 2002}

On November, 2002 V838 Mon was classified by Evans et al. (2002)
as an L-supergiant. The infrared specrum of V838 Mon shows deep
absorption bands of \HHO. In the optical spectra there are strong TiO bands 
as well as bands of a few diatomic molecules (Fig. \ref{__nov}).

A numerical analysis of V838 Mon spectrum was carried out by
Pavlenko et al. (2006). We showed that the slope of
the spectrum on 0.6 -- 1 micron depends on \Tef. The best fit was
obtained for \Tef = 2000 K (Fig. \ref{__nov}). 

It is worth noting that the Balmer hydrogen lines, from H$_{\beta}$ to                                                      
the Balmer jump, are clearly seen in the observed B3V spectrum. 
Using the 
Kurucz (1993) model atmospheres we computed theoretical spectra using WITA6 
for different \Tef\ and $\log g$.
The best fit to the observed spectrum provides   
\Tef=18000 - 23000~K  and  $\log g$ = 4.0 - 4.5.

Our fits of the combined (supergiant+dwarf) synthetic spectrum  to the 
observed spectrum provide the unique possibility for 
the determination of the radius of the V838
Mon photosphere.
This is only possible if V838 Mon and the B3V dwarf form a physically 
bound system.

Indeed, the radius and effective temperature of the ``normal'' 
B3V star should be 
 $\sim$ 6 \Rs and $\sim$ 20~000 K, respectively. 
Then, the ratio of the theoretical fluxes around 7000\AA~ computed by 
WITA6 for the model atmospheres of  
different effective temperatures is $\sim$ 100.
Using one simple
suggestion that this ratio can be applied to integrated fluxes
from both stars, we get for November 2002

$$(R_{V838 Mon}/R_{B3V})^2 = 100*(T_{\rm eff}^{B3V}/T_{\rm eff}^{V838 Mon})^4$$
and  $$R_{V838 Mon} = 6000 R_{\odot}.$$

Again, we note that this estimation is only correct if V838 Mon and the B3V
dwarf are located at the same distance from the Sun.

\section{Conclusion}

From February until November 2002, we do not see any chemical
pecularities in the spectrum of V838 Mon. The observed spectrum
resembles the spectrum of the photosphere of a normal supergiant
of slightly reduced metallicity [Fe/H] = -0.4 or -0.2 (see Kipper et
al. 2004, Kaminsky \& Pavlenko, 2005). It therefore provides indirect
observation of the evolution of a massive star or binary 
system with a massive component.

Then, V838 Mon represents a case of very slow evolution. The
spectral sequence of V838 Mon can be fitted by the sets of
computations for model atmospheres with hydrostatic equilibrium.
Weak or intermediate strong spectral lines form in the atmospheres
with decreasing temperature outwards. Only the strongest lines in
February-March 2002 show P Cyg profiles. The relaxation time after
the stellar flash(es) does not exceed 10 days. All these factors
should be accounted for in future theoretical models.

\acknowledgements

YP thanks SOC \& LOC of the Meeting for the invitation
and support of his participation. YP's participation was 
partly supported also by R.Rebolo
and E.Martin (IAC).
We are grateful to our colleagues
Nye Evans, Jacco van Loon (Keele University), 
Lisa Crause (SAAO), V.Klochkova(SAO), T.Kipper(Tartu), U.Munari(Padova)
for the excellent collaboration. We thank Gregory J. Harris (UCL) for some remarks on the text.  We thank the authors of the Arcturus atlas for making it 
available through ftp.



\question{Hirschi} With Teff=2000K, a radius of the order of 1000 R seems
          reasonable at least for Red Super Giants

\answer{Pavlenko} OK, we get the radius 6000 R$_{\odot}$.

\question{Goranskij} In the spectra shown in web page of Conference,
            we see that details of L supergiant expand more and more
            in the blue region. This means that the star becomes hotter
            and hotter. And this tendency is confirmed by photometry. 
            But your calculations show cooling of the star. Why?

\answer{Pavlenko} IR region is better to find temperature of so cool star, because
           there is the maximum of flux located. Note, 
	   by definition the blue part of the 
	   spectra should show stronger dependence on shock waves or 
	   any other processes of stellar activity. 
	   Fortunately, there is only small part 
	   of the total/integrated flux of our cool supergiant is located.


\begin{thebibliography}{}
\bibitem[]{} Barber R.J., Tennyson J., Harris G.J., Tolchenov R., 2006, MNRAS, 
368, 107
\bibitem[Burrows et al. (2002)]{burrows02}{Burrows, A., Ram, S. R., Bernath,
         P., 2002, ApJ, 577, 986}
\bibitem[Crause et al. (2003)]{crause2003}{Crause, L.A., et al. 2003,
         MNRAS, 341, 785}
\bibitem[Desidera \& Munari (2002)]{desidera2002}{Desidera, S., Munari, U.,
         2002, IAUC7982}
\bibitem[Dulick et al. (2003)]{dulick03}{Dulick, M., Bauschlincher, C. W.,
         Burrows, A. 2003, ApJ, 594, 651}
\bibitem[Evans et al. (2003)]{evans03}{Evans, A., Geballe, T. R., Rushton, M.
         T., Smalley, B.,  van Loon, J. Th., Eyres, S. P. S., Tyne, V. H., 2003,
     MNRAS, 343, 1054}
\bibitem[Hauschildt et al. (1999)]{hauschildt99}{Hauschildt, P. H., Allard, F.,
         Baron, E., 1999, ApJ, 512, 377}
\bibitem[]{} Hinkle K., Wallace L., Valenti J., Harmer D. Visible and near IR Atlas of the 
Arcturus spectrum 3727 -- 9300 \AA.
ftp://ftp.noao.edu/catalogs/arcturusatlas/visual/
\bibitem[Kaminsky \& Pavlenko (1995)]{kaminsky05}{Kaminsky, B. M., Pavlenko, Y.
         P., 2005, MNRAS, 357, 38}
\bibitem[Kipper et al. (2004)]{Kipper2004}{Kipper, T., Klochkova, V.G.,
         Annuk, K., et al. 2004, A\&A, 416, 1107}
\bibitem[Kupka et al. (1999)]{kupka99}{Kupka, F., Piskunov, N., Ryabchikova, T.
         A., Stempels, H. C., Weiss, W. W. 1999, A\&AS, 138, 119}
\bibitem[Kurucz(1993)]{kurucz93}{Kurucz, R. L., 1993, CDROMs 1-22,
         Harvard-Smisthonian Astonomical Observatory}
\bibitem[van Loon et al. (2005)]{loon2005}{van Loon, J. Th., Evans, A., Rushton,
M.       T., Smalley, B. 2005, A\&A, V, P}
\bibitem[Munari et al. (2002)]{Munari2002}{Munari, U., Henden, A., Kiyota,
         S., et al. 2002, AA, 389, L51}
\bibitem[Munari et al. (2005)]{munari05}{Munari, U., et al., 2005, A\&A, 434,
         1107}
\bibitem[Pavlenko (2000)]{pavlenko00}{Pavlenko, Ya. 2000, Astron. Rept., 44, 219}
\bibitem[Pavlenko et al. (2006)]{p2006}
Ya. V. Pavlenko, J. Th. van Loon, A. Evans et al. 2006, A\&A, accepted 
(astro-ph/0609225)
\bibitem[Plez (1998)]{plez98}{Plez, B., 1998, A\&A, 337, 495}
\bibitem[Rushton et al. (2005)]{rushton05}{Rushton, M. T., Geballe, T. R.,
        Filippenko, A. V., Chornock, R., Li, W., Leonard, D. C., Foley, R. J.,
        Evans, A., Smalley, B., van Loon, J. Th., Eyres, S. P. S., 2005b,
        MNRAS, 360, 1281}
\bibitem[Tylenda (2005)]{tylenda05}{Tylenda, R., 2005, A\&A, 436, 1009}
\bibitem[van Loon et al. (2004)]{loon2004}{van Loon, J.Th., Evans, A., Rushton,
         M.T, Smalley, B. 2004, AA, 427, 193}
\end{thebibliography}
\end{document}